\documentclass[10pt,english,aps,showkeys,showpacs,nofootinbib]{revtex4-1}
\usepackage[T1]{fontenc}
\usepackage{textcomp}
\usepackage[utf8]{inputenc}
\usepackage{amsmath}
\usepackage{amssymb}
\usepackage{stackrel}

\makeatletter
\usepackage[utf8]{inputenc}
\usepackage{todonotes}
\usepackage[bottom]{footmisc}
\usepackage{graphicx}
\usepackage{dcolumn}
\usepackage{bm}
\usepackage{todonotes}
\usepackage{hyperref}
\usepackage[scale=0.75]{geometry}
\makeatother

\usepackage{babel}
\begin{document}
\selectlanguage{english}
\title{On The Feshbach–Villars Oscillator (FVO) Under Coulomb-Type Potential
In The Cosmic Dislocation Space-Time}
\author{Abdelmalek Bouzenada }
\email{abdelmalek.bouzenada@univ-tebessa.dz ; abdelmalekbouzenada@gmail.com}

\affiliation{Laboratoire de Physique Appliquée et Théorique~\\
 Université Larbi-Tébessi-, Tébessa, Algeria}
\author{Abdelmalek Boumali}
\email{boumali.abdelmalek@gmail.com}

\affiliation{Laboratoire de Physique Appliquée et Théorique~\\
 Université Larbi-Tébessi-, Tébessa, Algeria}
\author{Marwan Al-Raeei}
\email{mhdm-ra@scs-net.org ; mn41@liv.com}

\affiliation{Faculty of Science,Damascus~\\
 University,Damascus,Syria}
\date{\today}
\begin{abstract}
In this paper, we investigate the quantum mechanical dynamics of the
massive and relativistic Feshbach-Villars oscillator in cosmic dislocation
space-time induced by a coulomb-type potential. The first-order Feshbach-Villars
version of the Klein-Gordon equation is used to find movement equations.
Wave functions and associated energy have been calculated (both in
the free case and in the interaction case).

We analyze the impact of dislocation topology on this interaction.
As a result, the effect of the dislocation on the quantum system under
study is examined.
\end{abstract}
\maketitle

\section{Introduction }

The impact of the gravitational field on the dynamics of quantum mechanical
systems is of wide interest. On the one hand, Einstein's theory of
general relativity (GR)\citep{key-56} gives a convincing explanation
of gravity as a geometric characteristic of space-time. It proves,
in particular, that the classical gravitational field is a manifestation
of space-time curvature. It has predicted the existence of gravitational
waves \Citep{key-57} and black holes \Citep{key-58} for example.Quantum
mechanics (QM) is the framework for understanding the behavior of
particles on a tiny scale \Citep{key-59}. It is a very effective
theory (usually quantum field theory) in describing how small particles
interact and how three of nature's four fundamental forces emerge:
weak, strong, and electromagnetic interactions \Citep{key-60}. However,
attempts to develop a unified theory that can reconcile general relativity
with quantum mechanics, i.e. a theory of quantum gravity, have hit
various roadblocks and technical challenges that have yet to be overcome
\Citep{key-6,key-7} .

One basic method for establishing a broad picture of how the gravitational
field affects relativistic particles at the quantum level is to generalize
aspects of the relativistic dynamics of particles in flat Minkowski
space to an arbitrary curved background geometry\citep{key-8,key-9}.
As a result, the approach can be adapted to deal with different models
in which the concept of curvature appears, incorporating more predictions
on the values of macroscopic observables that are required to make
relevant experimental verification of certain phenomenological consequences,
particularly in astrophysics and cosmology.Furthermore, comprehending
the thermodynamic behavior of relativistic particles where gravitational
effects must be considered \citep{key-10,key-11,key-12}, as well
as analyzing the associated features, i.e., the fundamental statistical
quantities, would provide the possibility of obtaining useful and
essential results in the context of describing the quantum behavior
of gravity.

Topological defects (domain walls, cosmic strings, monopoles, and
textures) have been extensively researched over the last few decades
and continue to be one of the most active disciplines in condensed
matter physics, cosmology, astrophysics, and elementary particle models.
It is thought that these structures arose as a result of the Kibble
mechanism \citep{key-68}. \Citep{key-14,key-15}, where the defects
emerge during the cooling of the early universe in symmetry-breaking
phase transitions \citep{key-16,key-17}. The particular fault in
question is cosmic strings (for additional information, see \Citep*{key-73}).
These items (whether static or revolving) can have discernible impacts.
They, for example, offer a potential technique for seeding galaxy
formation and gravitational lensing effects.Furthermore, by investigating
cosmic strings and their characteristics, we may learn a lot about
particle physics at very high energies in many settings. Furthermore,
the potential that cosmic strings may act like superconducting wires
has been proposed in current physics, with fascinating implications.

The harmonic oscillator (HO) has long been recognized as an important
instrument in many branches of theoretical physics \Citep{key-19}.
It is a well-studied perfectly solvable model that may be used to
examine different difficult issues using quantum mechanics \Citep{key-74}.
Furthermore, the relativistic extension of the quantum harmonic oscillator
gives a useful model for understanding a wide range of molecular,
atomic, and nuclear interactions. Indeed, when working with such a
model, the property of having a complete set of precise analytical
solutions can give rise to radically different interpretations of
many mathematical and physical events, and hence related applications
can be obtained via the underlying formulation.

The behavior of various relativistic quantum systems is critically
dependent on the Dirac oscillator (DO). As It\^{o} et al. \citep{key-75}
indicated in previous advances of spin-1/2 particle dynamics with
a linear trajectory. They demonstrated that the system's non-relativistic
limit yields an ordinary harmonic oscillator with a large spin-orbit
coupling term. Actually, Moshinsky and Szczepaniak \citep{key-76}
found that the above-mentioned DO could be derived from the free Dirac
equation by introducing an external linear potential through a minimum
replacement of the momentum operator $\hat{p}\longrightarrow\hat{p}-im\omega\beta\hat{r}$.It
is worth noting that, in addition to the theoretical focus on researching
the DO, substantial insights may be achieved by examining physical
interpretation, which is undoubtedly important in comprehending many
pertinent applications.

Inspired by DO, a similar formalism for the case of bosonic particles
was developed, and it was dubbed a Klein-Gordon oscillator (KGO) \citep{key-23,key-24}.
Several writers have lately been working on the covariant version
of this model in curved space-times and other configurations. Numerous
contributions have been made to the subject of relativistic quantum
motions of scalar and vector particles under gravitational effects
produced by various curved space-time geometries; for example, Ref.
\Citep{key-79} studies the problem of the interaction between KGO
coupled harmonically with topological defects in Kaluza-Klein theory.
Ref. \Citep{key-80} investigates the relativistic quantum dynamics
of spin-0 particles in a spinning cosmic string space-time with Coulomb-type
scalar and vector potentials. Furthermore, rotating effects on the
scalar field in cosmic string space-time, space-time with space-like
dislocation, and space-time with spiral dislocation have been examined
in Ref. \Citep{key-81}. Recently, the authors of Ref. \Citep{key-28}
examined the KGO in a cosmic string space-time and investigated the
effects of the rotating frame and non-commutativity in momentum space.
Furthermore, in Ref. \Citep{key-82}, the KGO was exposed to a magnetic
quantum flux in the presence of a Cornell-type scalar and Coulomb-type
vector potentials in a spinning cosmic string space-time.

Attempts to investigate the relativistic spin-0, spin-1 bosons and
spin-1/2 fermions wave functions and their time evolution have been
pursued by various authors \Citep{key-30,key-31,key-32} making use
of the Hamiltonian form i.e, having Schrodinger-type equations. The
so-called Feshbach-Villars (FV) equations \Citep{key-33} are of particular
interest in this respect. These equations were initially constructed
in the purpose of permitting a relativistic single particle interpretation
of the second-order KG equation. For the later case, FV equations
originate from splitting the KG wave function into two components
in order to obtain an equation with first order time derivative. In
recent decades, a number of papers have been produced with the aim
of exploring the relativistic dynamical properties of single particles
and solving their wave equations by adopting the FV scheme (e.g,Refs.\Citep{key-34,key-35,key-36,key-37,key-38,key-39,key-40}
and other related references cited therein),Bouzenada et al \citep{key-41}
investigate the Feshbach-Villars oscillator (FVO) case in spinning
cosmic string space-time and discuss some findings ( thermal properties
and density of this systhem).

This paper investigates the effect of a Coulomb-type potential on
the Klein-Gordon oscillator. Several publications\citep{key-42,key-43,key-44,key-45,key-46}
have recently considered the confinement of a relativistic scalar
particle to a Coulomb potential. The approach for inserting a scalar
potential into the Klein-Gordon equation is as follows, as explained
in Ref. \citep{key-45}. The electromagnetic 4-vector potential is
introduced in the same way.This is accomplished by altering the momentum
operator $p_{\alpha}=i\partial_{\alpha}$ as follows:$p_{\alpha}\rightarrow p_{\alpha}-qA_{\alpha}\left(x\right)$.
Another technique was provided in Ref. \citep{key-46} by changing
the mass term to:$m\rightarrow m+S(\overrightarrow{r},t)$ , where
$S(\overrightarrow{r},t)$ denotes the scalar potential.This change
in the mass term has been studied in recent decades, for example,
by examining the behavior of a Dirac particle in the presence of a
static scalar potential and a Coulomb potential \citep{key-46} and
a relativistic scalar particle in cosmic string spacetime \citep{key-47}.
We study the effect of a Coulomb-type potential on the Klein-Gordon
oscillator in this work by introducing the scalar potential as a modification
to the mass component in the Klein-Gordon equation.We obtain bound
state solutions to the Klein-Gordon equation for both attractive and
repulsive Coulomb-type potentials and demonstrate a quantum effect
characterized by the angular frequency of the Klein-Gordon oscillator
being dependent on the quantum numbers of the system, implying that
not all angular frequency values are allowed.

This paper is organized as follows. In the next section, we derive
the FV equations for scalar bosons in Mikowski and static Cosmic string
space-time, taking both the free and interaction cases into account.
We present the KG oscillator in Hamiltonian form, then solve the resulting
equations to get the eigenstates and energy levels. Section 3 extends
the preceding equations by considering the geometry of the cosmic
dislocation space-time; the same technique as Section 2 is applied
in this section. Section 4 investigates the Feshbach-Villars oscillator
and the quantification of the energy system based on these variables,we
are using Frobenius method. Section 5 contains our conclusions.We
shall always use natural units $\hbar=c=1$ throughout the article,
and our metric convention is $(+,-,-,-)$.

\section{The FV Representation of Feshbach-Villars (Spin-0) in Minkowski Space-time}

\subsection{An Overview of the Feshbach-Villars Approximation}

This section discusses the relativistic quantum description of a spin-0
particle propagating in Minkowski space-time using the metric tensor
$\eta_{\mu\nu}=\text{diag}\left(1,-1,-1,-1\right)$.The usual covariant
KG equation for a scalar massive particle $\Phi$ with mass $m>0$
is \Citep{key-48,key-49}
\begin{equation}
\left(\eta^{\mu\nu}D_{\mu}D_{\nu}+m^{2}\right)\Phi(x,t)=0,\label{eq:01}
\end{equation}
The minimally-coupled covariant derivative is denoted by $D_{\mu}=\left(p_{\mu}-ieA_{\mu}\right)$.
The classical four momentum is $p_{\mu}=\left(E,-p_{i}\right)$, while
the electromagnetic four potential is $A_{\mu}=\left(A_{0},-A_{i}\right)$.
The magnitude of the particle charge is given by e.

It is worth noting at this point that \eqref{eq:01} may be expressed
in Hamiltonian form with the time first derivative, i.e. as a Schrödinger-type
equation.
\begin{equation}
\mathcal{H}\Phi\left(x,t\right)=i\,\frac{\partial}{\partial t}\Phi(x,t),\label{eq:02}
\end{equation}
The Hamiltonian $\mathcal{H}$ may be defined using the FV linearization
process, which involves converting \ref{eq:01} to a first order in
time differential equation. The two component wave function \Citep{key-46}
is introduced. \Citep{key-50},

\begin{equation}
\Phi(x,t)=\left(\begin{array}{c}
\phi_{1}(x,t)\\
\phi_{2}(x,t)
\end{array}\right)=\frac{1}{\sqrt{2}}\left(\begin{array}{c}
1+\frac{i}{m}\mathcal{D}\\
1-\frac{i}{m}\mathcal{D}
\end{array}\right)\psi(x,t),\label{eq:03}
\end{equation}
Here, $\psi(x,t)$ obeys the KG wave equation, and $\mathcal{D}$
is defined in such a way that
\begin{equation}
\mathcal{D}=\frac{\partial}{\partial t}+ieA_{0}(x),\label{eq:04}
\end{equation}
The aforementioned transformation \eqref{eq:03} involves inserting
wave functions that meet the requirements.
\begin{equation}
\psi=\phi_{1}+\phi_{2},\qquad i\mathcal{D}\psi=m\left(\phi_{1}-\phi_{2}\right).\label{eq:05}
\end{equation}
It is more convenient to write for our subsequent review,
\begin{equation}
\begin{aligned}\phi_{1} & =\frac{1}{2m}\left[m+i\frac{\partial}{\partial t}-eA_{0}\right]\psi\\
\phi_{2} & =\frac{1}{2m}\left[m-i\frac{\partial}{\partial t}+eA_{0}\right]\psi,
\end{aligned}
\label{eq:06}
\end{equation}
Eq. \eqref{eq:01} becomes equivalent
\begin{equation}
\begin{aligned}\left[i\frac{\partial}{\partial t}-eA_{0}\right]\left(\phi_{1}+\phi_{2}\right) & =m\left(\phi_{1}-\phi_{2}\right)\\
\left[i\frac{\partial}{\partial t}-eA_{0}\right]\left(\phi_{1}-\phi_{2}\right) & =\left[\frac{\left(p_{i}-eA_{i}\right)^{2}}{m}+m\right]\left(\phi_{1}+\phi_{2}\right),
\end{aligned}
\label{eq:07}
\end{equation}
The addition and subtraction of these two equations yields a system
of first order coupled differential equations
\begin{equation}
\begin{aligned}\frac{\left(p_{i}-eA_{i}\right)^{2}}{2m}\left(\phi_{1}+\phi_{2}\right)+\left(m+eA_{0}\right)\phi_{1} & =i\frac{\partial\phi_{1}}{\partial t}\\
\frac{-\left(p_{i}-eA_{i}\right)^{2}}{2m}\left(\phi_{1}+\phi_{2}\right)-\left(m-eA_{0}\right)\phi_{2} & =i\frac{\partial\phi_{2}}{\partial t},
\end{aligned}
\label{eq:08}
\end{equation}
The FV Hamiltonian of a scalar particle in the presence of electromagnetic
interaction may be expressed using Eqs. \eqref{eq:08} as

\begin{equation}
\mathcal{H}_{K-G}=\left(\tau_{3}+i\tau_{2}\right)\frac{\left(p_{i}-eA_{i}\right)^{2}}{2m}+m\tau_{3}+eA_{0}(x),\label{eq:09}
\end{equation}
where $\tau_{i}\,\left(i=1,2,3\right)$ are the conventional $2\times2$
Pauli matrices given by
\begin{equation}
\tau_{1}=\left(\begin{array}{cc}
0 & 1\\
1 & 0
\end{array}\right),\quad\tau_{2}=\left(\begin{array}{cc}
0 & -i\\
i & 0
\end{array}\right),\quad\tau_{3}=\left(\begin{array}{cc}
1 & 0\\
0 & -1
\end{array}\right).
\end{equation}
It's worth noting that the Hamiltonian \eqref{eq:09} meets the generalized
hermicity requirement (If there is an invertible, Hermitian, linear
operator $\beta$ such that $\mathcal{H}^{\dagger}=\beta\mathcal{H}\beta^{-1}$,
the Hamiltonian $\mathcal{H}$ is said to be pseudo-Hermitian. \Citep{key-51}).

\begin{equation}
\mathcal{H}_{K-G}=\tau_{3}\left(\mathcal{H}_{K-G}^{\dagger}\right)\tau_{3},\qquad\mathcal{H}_{K-G}^{\dagger}=\tau_{3}\left(\mathcal{H}_{K-G}\right)\tau_{3}.
\end{equation}
The one dimensional FV Hamiltonian reduces to for free particle propagation,
i.e., no interaction is assumed left $\left(A_{\mu}=0\right)$.
\begin{equation}
\mathcal{H}_{0}=\left(\tau_{3}+i\tau_{2}\right)\frac{p_{x}^{2}}{2m}+m\tau_{3},\label{eq:12}
\end{equation}
The solutions to the time-independent free Hamiltonian are simply
stationary states. Assuming the solution \Citep{key-36},
\begin{equation}
\Phi\left(x,t\right)=\Phi\left(x\right)e^{-iEt}=\left(\begin{array}{c}
\phi_{1}\left(x\right)\\
\phi_{2}\left(x\right)
\end{array}\right)e^{-iEt},\label{eq:13}
\end{equation}
with E denoting the system's energy. As a result, Eq. \eqref{eq:02}
may be represented as
\begin{equation}
\mathcal{H}_{0}\Phi\left(x\right)=E\Phi\left(x\right),\label{eq:14}
\end{equation}
This is the one-dimensional FV equation of the free relativistic spin-0
particle, and it is performed in order to have an alternate Schrödinger-type
to KG equation. In what follows, the aforesaid approach will be utilized
to determine the dislocation solutions to wave equations in curved
space-time.

\section{The FV Representation of Spin-0 Particle in Cosmic Dislocation Space-time}

The goal of this part is to investigate the KGO in the backdrop geometry
of a cosmic string using the FV technique. It is widely known that
the generally covariant relativistic wave equations of a scalar particle
in a Riemannian space-time characterized by the metric tensor $g_{\mu\nu}$
may be found by reformulating the KG equation so that( see, e.g, the
textbooks\citep{key-8,key-9} )
\begin{equation}
\left(\square+m^{2}-\mathcal{\xi}R\right)\Phi(x,t)=0,\label{eq:15}
\end{equation}
where $\square$ is the Laplace-Beltrami operator denoted by
\begin{equation}
\square=g^{\mu\nu}D_{\mu}D_{\nu}=\frac{1}{\sqrt{-g}}\partial_{\mu}\left(\sqrt{-g}g^{\mu\nu}\partial_{\nu}\right),\label{eq:16}
\end{equation}
$\xi$ denotes a real dimensionless coupling constant, and R is the
Ricci scalar curvature given by $R=g^{\mu\nu}R_{\mu\nu}$, where $R_{\mu\nu}$
is the Ricci curvature tensor. The inverse metric tensor is $g^{\mu\nu}$,
and $g=\det\left(g^{\mu\nu}\right)$.

We would now want to investigate the quantum dynamics of spin-0 particles
in the space-time caused by a (3+1)-dimensional dislocation, as well
as develop the relevant FV formulation. 

\subsection{Feshbach-Villars oscillator in the cosmic dislocation space-time}

Before we study the KGO in the Hamiltonian representation, let us
first derive the KG wave equation for the free relativistic scalar
particle propagating in the cosmic string space-time that is assumed
to be static and cylindrically symmetric.

In this part, we will discuss the topological defect that serves as
the foundation for our study. Inspired by the description of an edge
dislocation in crystalline solids, we build a generalization of this
topological defect in gravity. We can observe that an edge dislocation
is a spiral dislocation, which is a deformation of a circle into a
spiral \citep{key-52}. The line element describing the space-time
backdrop with this topological defect is (using the units $\hbar=c=1$)
\citep{key-53,key-54,key-55} 
\begin{align}
ds^{2} & =g_{\mu\nu}dx^{\mu}dx^{\nu}\nonumber \\
 & =dt^{2}-dr^{2}-\left(\alpha^{2}r^{2}+\chi^{2}\right)d\varphi^{2}+2\chi dzd\varphi-dz,\label{eq:17}
\end{align}
where $\chi$ is a constant value relating to the defect's distortion.
By $\chi=\frac{\left|\overrightarrow{b}\right|}{2\pi}$, the parameter
is also connected to the Burgers vector $\overrightarrow{b}$. Here
$-\infty\le t\le+\infty$, $r\ge0$, $0\le\varphi\le2\pi$, $-\infty\le z\le+\infty$,and
$\alpha\in[0,1[\:$: is the angular parameter that defines the angular
deficit $\delta\varphi=2\pi(1-\alpha)$, which is connected to the
string's linear mass density mu by $\alpha=1-4\mu$(It should be noted
that this metric provides an accurate solution to Einstein's field
equations for $0\le\mu<1/4,$, and that by setting $\varphi^{\prime}=\alpha\varphi$,
it represents a flat conical outer space with angle deficit $\delta\phi=8\pi\mu$).

When the metric and inverse metric tensor components are, respectively,
\begin{equation}
g_{\mu\nu}=\left(\begin{array}{cccc}
1 & 0 & 0 & 0\\
0 & -1 & 0 & 0\\
0 & 0 & -\left(\alpha^{2}r^{2}+\chi^{2}\right) & \chi\\
0 & 0 & \chi & -1
\end{array}\right),\quad g^{\mu\nu}=\left(\begin{array}{cccc}
1 & 0 & 0 & 0\\
0 & -1 & 0 & 0\\
0 & 0 & -\frac{1}{\alpha^{2}r^{2}} & -\frac{\chi}{\alpha^{2}r^{2}}\\
0 & 0 & -\frac{\chi}{\alpha^{2}r^{2}} & -\left(1+\frac{\chi^{2}}{\alpha^{2}r^{2}}\right)
\end{array}\right)\label{eq:18}
\end{equation}
It is worth noting that the topic of spinless heavy particles in the
geometry formed by a cosmic dislocation background has been studied
in a number of studies (for example,\citep{key-56,key-57}).

To obtain the FV form of the KG wave equation in curved manifolds,
we will use the approach provided in references \citep{key-58,key-59}.
The generalized Feshbach-Villars transformation is used (GFVT) An
identical transformation for characterizing both large and massless
particles was presented earlier in Ref.\citep{key-60}. The components
of the wave function $\Phi$ in the GFVT are provided by \citep{key-58}.
\begin{equation}
\psi=\phi_{1}+\phi_{2},\qquad i\tilde{\mathcal{D}}\psi=\mathcal{N}\left(\phi_{1}-\phi_{2}\right),\label{eq:19}
\end{equation}
where $\mathcal{N}$ is an arbitrary nonzero real parameter and $\tilde{\mathcal{D}}=\frac{\partial}{\partial t}+\mathcal{Y},$
is specified, with
\begin{equation}
\mathcal{Y}=\frac{1}{2}\left\{ \partial_{i},\sqrt{-g}\frac{g^{0i}}{g^{00}}\right\} ,\label{eq:20}
\end{equation}
The anti-commutator is denoted by the curly bracket in Eq. \eqref{eq:20}.
The Hamiltonian for the aforementioned transformation is
\begin{equation}
\mathcal{H}_{GFVT}=\tau_{z}\left(\frac{\mathcal{N}^{2}+\mathcal{T}}{2\mathcal{N}}\right)+i\tau_{y}\left(\frac{-\mathcal{N}^{2}+\mathcal{T}}{2\mathcal{N}}\right)-i\mathcal{Y},\label{eq:21}
\end{equation}
with
\begin{align}
\mathcal{T} & =\partial_{i}\frac{G^{ij}}{g^{00}}\partial_{j}+\frac{m^{2}-\xi R}{g^{00}}+\frac{1}{\mathcal{F}}\nabla_{i}\left(\sqrt{-g}G^{ij}\right)\nabla_{j}\left(\frac{1}{\mathcal{F}}\right)+\sqrt{\frac{\sqrt{-g}}{g^{00}}}G^{ij}\nabla_{i}\nabla_{j}\left(\frac{1}{\mathcal{F}}\right)+\frac{1}{4\mathcal{F}^{4}}\left[\nabla_{i}\left(\mathcal{U}^{i}\right)\right]^{2}\nonumber \\
 & \qquad-\frac{1}{2\mathcal{F}^{2}}\nabla_{i}\left(\frac{g^{0i}}{g^{00}}\right)\nabla_{j}\left(\mathcal{U}^{i}\right)-\frac{g^{0i}}{2g^{00}\mathcal{F}^{2}}\nabla_{i}\nabla_{j}\left(\mathcal{U}^{i}\right),\label{eq:22}
\end{align}
where
\begin{equation}
G^{ij}=g^{ij}-\frac{g^{0i}g^{0j}}{g^{00}},\qquad\mathcal{F}=\sqrt{g^{00}\sqrt{-g}},\qquad\mathcal{U}^{i}=\sqrt{-g}g^{0i}.\label{eq:23}
\end{equation}
Here and everywhere $(i,j=1,2,3)$. We see that the initial FV transformations
are fulfilled for $\mathcal{N}=m$.

Now, using the metric \ref{eq:25}, it is simple to see that $R=0$,
implying that space-time is locally flat (no local gravity), and so
the coupling component is vanishing. The condition $\xi=0$ is known
as minimum coupling. However, in massless theory, $\xi$ equals 1/6.
(in 4 dimensions). The equations of motion are then conformally invariant
in this later instance.

A simple computation yields $\mathcal{Y}=0,$ and we then obtain
\begin{equation}
\mathcal{T}=-\frac{d^{2}}{dr^{2}}-\frac{1}{r}\frac{d}{dr}-\frac{1}{\alpha^{2}r^{2}}\frac{d^{2}}{d\varphi^{2}}-\left(1+\frac{\chi^{2}}{\alpha^{2}r^{2}}\right)\frac{d^{2}}{dz^{2}}-\frac{2\chi}{\alpha^{2}r^{2}}\left(\frac{d}{d\varphi}\frac{d}{dz}\right)+m^{2},\label{eq:24}
\end{equation}
Using these techniques to obtain the Hamiltonian \ref{eq:21}, one
may suppose a solution of the type we want cylindrically symmetric,
i.e. solutions with rotational symmetry in the (x,y)-plane, due to
the temporal and angular independence in the metric \ref{eq:25}.
\begin{equation}
\Phi(t,r,\varphi,z)=\Phi(r)e^{-i\left(Et-j\varphi-Kz\right)},\label{eq:25}
\end{equation}
where $j=0,\pm1,\pm2,..$ are the eigenvalues of the $z$-component
of the angular momentum operator. The KG equation \ref{eq:15} may
be written equivalently to the following two coupled equations
\begin{align}
\left(\mathcal{N}^{2}+\mathcal{T}\right)\phi_{1}+\left(-\mathcal{N}^{2}+\mathcal{T}\right)\phi_{2} & =2\mathcal{N}E\phi_{1}\nonumber \\
-\left(\mathcal{N}^{2}+\mathcal{T}\right)\phi_{2}-\left(-\mathcal{N}^{2}+\mathcal{T}\right)\phi_{1} & =2\mathcal{N}E\phi_{2},\label{eq:26}
\end{align}
The sum and difference of the two previous equations yields a second
order differential equation for the field $\psi$. As a result, the
radial equation is as follows:
\begin{equation}
\left[\frac{d^{2}}{dr^{2}}+\frac{1}{r}\frac{d}{dr}-\frac{\zeta^{2}}{r^{2}}+\kappa\right]\psi\left(r\right)=0,\label{eq:27}
\end{equation}
where we have set
\begin{equation}
\zeta=\frac{j+\chi K}{\alpha},\qquad\kappa=\sqrt{E^{2}-m^{2}-K^{2}}.\label{eq:28}
\end{equation}
We can observe that Eq. \eqref{eq:27} is a Bessel equation and its
general solution is defined by \citep{key-61}
\begin{equation}
\psi\left(r\right)=A\,J_{|\zeta|}\left(\kappa r\right)+B\,Y_{|\zeta|}\left(\kappa r\right),\label{eq:29}
\end{equation}
where $J_{|\zeta|}\left(\kappa r\right)$ and $Y_{|\zeta|}\left(\kappa r\right)$
are the Bessel functions of order $\zeta$ and of the first and the
second kind, respectivement. Here $A$ and $B$ are arbitrary constants.
We notice that at the origin when $\zeta=0$, the function $J_{|\zeta|}\left(\kappa r\right)\ne0$.
However, $Y_{|\zeta|}\left(\kappa r\right)$ is always divergent at
the origin. In this case, we will consider only $J_{|\zeta|}\left(\kappa r\right)$
when $\zeta\ne0$. Hence, we write the solution to Eq. \eqref{eq:27}
as follows
\begin{equation}
\psi\left(r\right)=A\,J_{\frac{|j+\chi K|}{\alpha}}\left(\sqrt{E^{2}-m^{2}-K^{2}}\,r\right),\label{eq:30}
\end{equation}
We can now express the whole two-component wavefunction of the spinless
heavy KG particle in the space-time of a cosmic dislocation using
this solution.
\begin{equation}
\psi\left(t,r,\varphi,z\right)=\left|\mathcal{C}_{1}\right|\left(\begin{array}{c}
1+\frac{E}{\mathcal{N}}\\
1-\frac{E}{\mathcal{N}}
\end{array}\right)e^{-i\left(Et-j\varphi-Kz\right)}\,J_{\frac{|j+\chi K|}{\alpha}}\left(\sqrt{E^{2}-m^{2}-K^{2}}\,r\right),\label{eq:31}
\end{equation}
The constant $\left|\mathcal{C}_{1}\right|$ can be obtained by applying
the appropriate normalization condition to the KG equation (e.g.,
see Ref.\citep{key-62,key-63}), but it is fortunate that failing
to determine the normalization constants throughout this manuscript
has no effect on the final results.

Now we'll look at the specific instance where we wish to extend the
GFVT for the KGO. In general, we must substitute the momentum operator
in Eq. \eqref{eq:15}. As a result, Eq. \eqref{eq:30} may be rewritten
as follows.
\begin{equation}
\mathcal{T}=\frac{1}{\sqrt{-g}}\left(\frac{\partial}{\partial r}-m\omega r\right)\left(-\sqrt{-g}\right)\left(\frac{\partial}{\partial r}+m\omega r\right)-\frac{1}{\alpha^{2}r^{2}}\frac{\partial^{2}}{\partial\varphi^{2}}-\left(1+\frac{\chi^{2}}{\alpha^{2}r^{2}}\right)\frac{d^{2}}{dz^{2}}-\frac{2\chi}{\alpha^{2}r^{2}}\left(\frac{d}{d\varphi}\frac{d}{dz}\right)+m^{2},\label{eq:32}
\end{equation}
Similarly, the following differential equation may be obtained using
a simple calculation based on the approach described above.
\begin{equation}
\left[\frac{d^{2}}{dr^{2}}+\frac{1}{r}\frac{d}{dr}+m^{2}\omega^{2}r^{2}-\frac{\sigma^{2}}{r^{2}}+\delta\right]\psi\left(r\right)=0,\label{eq:33}
\end{equation}
with
\begin{equation}
\sigma^{2}=\left(\frac{j+\chi K}{\alpha}\right)^{2},\qquad\delta=E^{2}-m^{2}-K^{2}+2m\omega.\label{eq:34}
\end{equation}
The KGO for a spin-0 particle in the space-time of a static cosmic
string is given by Eq. \eqref{eq:33}. To derive the solution to this
problem, we first suggest a radial coordinate transformation.
\begin{equation}
\mathcal{Q}=m\omega r^{2},\label{eq:35}
\end{equation}
subsitutuing the expression for $\chi$ into Eq. \eqref{eq:33}, we
obtain
\begin{equation}
\left[\frac{d^{2}}{d\mathcal{Q}^{2}}+\frac{1}{\mathcal{Q}}\frac{\partial}{d\mathcal{Q}}-\frac{\sigma^{2}}{4\mathcal{Q}^{2}}+\frac{\delta}{4m\omega\mathcal{Q}}-\frac{1}{4}\right]\psi\left(\chi\right)=0.\label{eq:36}
\end{equation}
So, if we look at the asymptotic behavior of the wave function at
the origin and infinity, and we're looking for regular solutions,
we may assume a solution of the type
\begin{equation}
\psi\left(\mathcal{Q}\right)=\mathcal{Q}^{\frac{\left|\sigma\right|}{2}}e^{-\frac{\mathcal{Q}}{2}}F\left(\mathcal{Q}\right),\label{eq:37}
\end{equation}
As previously, we can plug this back into Eq. \eqref{eq:36}, and
we get
\begin{equation}
\mathcal{Q}\frac{d^{2}F\left(\mathcal{Q}\right)}{d\mathcal{Q}^{2}}+\left(|\sigma|+1-\chi\right)\frac{dF\left(\mathcal{Q}\right)}{d\mathcal{Q}}-\left(\frac{|\sigma|}{2}-\frac{\delta}{4m\omega}+\frac{1}{2}\right)F\left(\mathcal{Q}\right)=0,\label{eq:38}
\end{equation}
This is the confluent hypergeometric equation \citep{key-64}, the
solutions to which are defined in terms of the kind of confluent hypergeometric
function.
\begin{equation}
F\left(\mathcal{Q}\right)=_{1}F_{1}\left(\frac{\left|\sigma\right|}{2}-\frac{\delta}{4m\omega}+\frac{1}{2},+1,\mathcal{Q}\right),\label{eq:39}
\end{equation}
We should note that the solution \eqref{eq:39} must be a polynomial
function of degree $n$. However, taking $n\rightarrow\infty$ imposes
a divergence issue. We can have a finite polynomial only if the factor
of the last term in Eq. \eqref{eq:38} is a negative integer, meaning,
\begin{equation}
\frac{\left|\sigma\right|}{2}-\frac{\delta}{4m\omega}+\frac{1}{2}=-n\qquad,n=0,1,2,\cdots.\label{eq:40}
\end{equation}
With this result and the parameters \eqref{eq:34}, we may derive
the quantized energy spectrum of KGO in the cosmic dislocation space-time,
and hence,
\begin{equation}
E\left(n,\alpha,j\right)=\pm\sqrt{4m\omega n+\frac{2m\omega\left|j+\chi K\right|}{\alpha}+m^{2}+K^{2}},\label{eq:41}
\end{equation}
We may notice that the energy relies clearly on the angular deficit
$\alpha$. In other words, because to the presence of the wedge angle,
the curvature of space-time that is impacted by the topological defect,
i.e., the cosmic string, would affect the relativistic dynamics of
the scalar particle by creating a gravitational field.

The corresponding wave function is given by
\begin{equation}
\psi\left(t,r,\varphi,z\right)=\left|\mathcal{C}_{2}\right|\left(m\omega r^{2}\right)^{\frac{\left|j+\chi K\right|}{2\alpha}}e^{-\frac{m\omega r^{2}}{2}}{}_{1}F_{1}\left(\frac{\left|j+\chi K\right|}{2\alpha}-\frac{\delta}{4m\omega}+\frac{1}{2},\frac{\left|j+\chi K\right|}{\alpha}+1,m\omega r^{2}\right),\label{eq:42}
\end{equation}
Thereafter, the general eigenfunctions are written as
\begin{equation}
\psi\left(t,r,\varphi,z\right)=\left|\mathcal{C}_{2}\right|\left(\begin{array}{c}
1+\frac{E}{\mathcal{N}}\\
1-\frac{E}{\mathcal{N}}
\end{array}\right)\left(m\omega r^{2}\right)^{\frac{\left|j+\chi K\right|}{2\alpha}}e^{-\frac{m\omega r^{2}}{2}}e^{-i\left(Et-j\varphi-Kz\right)}{}_{1}F_{1}\left(\frac{\left|j+\chi K\right|}{2\alpha}-\frac{\delta}{4m\omega}+\frac{1}{2},\frac{\left|j+\chi K\right|}{\alpha}+1,m\omega r^{2}\right),\label{eq:43}
\end{equation}
where $\left|\mathcal{C}_{2}\right|$ is the normalization constant.

\subsection{In the cosmic dislocation space-time, the Feshbach-Villars oscillator
Coulomb-Type Potentials}

In this part, we will look at the KGO in the context of a cosmic dislocation.
The equations of motion of a scalar particle can be obtained by examining
the GFVT, as in the instance examined in Sec.\ref{sec:3}. Numerous
writers investigated the quantum dynamics of relativistic particles
in the space-time of a cosmicdislocation, and a variety of models
were addressed. Mazur, for example, explored the quantum mechanical
features of heavy (or massless) particles in the gravitational field
of a spinning cosmic dislocation in a prior publication \Citep{key-65}.
He demonstrated that energy should be quantized when the string has
non-zero rotational momentum.Subsequently, Gerbert and Jackiw \Citep{key-66}
showed solutions to the KG and Dirac equations in the (2+1)-dimensional
space-time produced by a large point particle with arbitrary angular
momentum. The vacuum expectation value of the stress-energy tensor
for a massless scalar field conformally related to gravity was explored
in Ref. \Citep{key-67}. The authors of \Citep{key-66} investigated
the behavior of a quantum test particle meeting the Klein-Gordon equation
in a spinning cosmic string's space-time.

Additionally, it was demonstrated in Ref. \Citep{key-68} that the
extrema of the field's energy for specified angular and linear momenta
may be defined as spinning cosmic string solutions of $U(1)$ scalar
field theory with a cylindrically symmetric energy density. Furthermore,
in Ref. \Citep{key-69}, topological and geometrical phases owing
to the gravitational field of a cosmic string with mass and angular
momentum were explored.

The gravitational effects of rotating cosmic strings have recently
piqued the curiosity of researchers studying the dynamics and characteristics
of relativistic quantum particles. Vacuum fluctuations for a massless
scalar field surrounding a spinning cosmic string, for example, were
examined using a renormalization approach in Ref. \Citep{key-70}.
Similarly, in Ref. \citep{key-71}, the vacuum polarization of a scalar
field in the gravitational backdrop of a spinning cosmic thread was
examined. Additionally, the authors of Ref. \Citep{key-72} used a
completely relativistic technique to investigate the Landau levels
of a spinless heavy particle in the spacetime of a spinning cosmic
string.

Wang et al. \Citep{key-73} studied the KGO linked to a homogeneous
magnetic field in the backdrop of a revolving cosmic string. In addition,
Ref.\Citep{key-74} addressed the problem of a spinless relativistic
particle subjected to a uniform magnetic field in the spinning cosmic
string space-time. Furthermore, in Ref. \Citep{key-75}, the relativistic
quantum dynamics of a KG scalar field exposed to a Cornell potential
in spinning cosmic string space-time were reported. In addition, in
Ref. \Citep{key-76}, the relativistic scalar charged particle in
a spinning cosmic string space-time with Cornell-type potential and
Aharonov-Bohm effect was studied.

The topic examined in Sec. \ref{sec:3} is extended to a more generic
space-time with non-zero angular momentum in the following discussion.
In this paper, we investigate a heavy, relativistic spin-0 particle
whose wave-function is represented by Psi and satisfies the KG equation
\ref{eq:15} in the space-time caused by a (3+1)-dimensional stationary
cosmic dislocation described in Coulomb-Type Potentials$\left(S\left(r\right)=\frac{\lambda}{r}=\pm\frac{\left|\lambda\right|}{r}\right)$.
\begin{equation}
\left[\frac{d^{2}}{dr^{2}}+\frac{1}{r}\frac{d}{dr}-\frac{\left(\frac{j+\chi K}{\alpha}\right)^{2}}{r^{2}}+\left(E-\frac{\lambda}{r}\right)^{2}-m^{2}-K^{2}\right]\varphi\left(r\right)=0,\label{eq:44}
\end{equation}
where 
\begin{equation}
\left[\frac{d^{2}}{dr^{2}}+\frac{1}{r}\frac{d}{dr}-\left(\frac{\left(\frac{j+\chi K}{\alpha}\right)^{2}-\lambda^{2}}{r^{2}}\right)-\frac{2\lambda E}{r}+E-m^{2}-K^{2}\right]\varphi\left(r\right)=0,\label{eq:45}
\end{equation}
After simple algebraic manipulations we arrive at the following second
order differential equation for the radial function $\varphi(r)$
\begin{equation}
\varphi\left(r\right)=\left|\mathcal{C}_{3}\right|r^{-\frac{1}{2}}\mathrm{WhittakerM}(-\frac{E\lambda}{\sqrt{-E^{2}+K^{2}+m^{2}}},\frac{-\alpha^{2}\lambda^{2}+(K\chi+j)^{2}}{\alpha^{2}},2\sqrt{-E^{2}+K^{2}+m^{2}}\,r),\label{eq:46}
\end{equation}
By simplifying the relationship between Whitakar and confluent hypergeometric
function\citep{key-61}, yields{\scriptsize
\begin{equation}
\varphi\left(r\right)=\left|\mathcal{C}_{3}\right|r^{-\frac{1}{2}}\mathrm{e}^{-\left(\sqrt{-E^{2}+K^{2}+m^{2}}\right)r}\left(2r\left(\sqrt{-E^{2}+K^{2}+m^{2}}\right)\right)^{^{\frac{(-2\lambda^{2}+1)\alpha^{2}+2(K\chi+j)^{2}}{2\alpha^{2}}}}F\left(r\right)\label{eq:47}
\end{equation}
}where{\scriptsize
\begin{equation}
F\left(r\right)=1F1\left(\frac{\left\{ \left(-\lambda^{2}+\frac{1}{2}\right)\alpha^{2}+\left(K\chi+j)^{2}\right)\right\} \sqrt{-E^{2}+K^{2}+m^{2}}+E\lambda\alpha^{2}}{\left(\sqrt{-E^{2}+K^{2}+m^{2}}\right)\alpha^{2}},\frac{\left(-2\lambda^{2}+1\right)\alpha^{2}+2\left(K\chi+j\right){}^{2}}{\alpha^{2}},2r\left(\sqrt{-E^{2}+K^{2}+m^{2}}\right)\right)\label{eq:48}
\end{equation}
}We should note that the solution must be a polynomial function of
degree $n$. However, taking $n\rightarrow\infty$ imposes a divergence
issue. We can have a finite polynomial only if the factor of the last
term in Eq. is a negative integer, meaning,
\begin{equation}
\frac{\left[\left(-\lambda^{2}+\frac{1}{2}\right)\alpha^{2}+\left(K\chi+j)^{2}\right)\right]\sqrt{-E^{2}+K^{2}+m^{2}}+E\lambda\alpha^{2}}{\sqrt{-E^{2}+K^{2}+m^{2}}\alpha^{2}}=-n\qquad,n=0,1,2,\cdots.\label{eq:49}
\end{equation}
With this result and the parameters \eqref{eq:34}, we may derive
the quantized energy spectrum of KGO in the cosmic dislocation space-time,
and hence,{\scriptsize
\begin{equation}
E(n)=\pm\left(\frac{2\left(\left(2\lambda^{2}+2n+1\right)\alpha^{2}+2(K\chi+j)^{2}\right)\sqrt{\left[\left(\lambda^{4}+2\left(n+1\right)\lambda^{2}+\left(n+\frac{1}{2}\right){}^{2}\right)\alpha^{4}+2\left(K\chi+j\right){}^{2}\left(\lambda^{2}+n+\frac{1}{2}\right)\alpha^{2}+\left(K\chi+j\right){}^{4}\right]\left(K^{2}+m^{2}\right)}}{\left[4\lambda^{4}+8\left(n+1\right)\lambda^{2}+\left(2n+1\right){}^{2}\alpha^{4}+8\left(K\chi+j\right){}^{2}\left(\lambda^{2}+n+\frac{1}{2}\right)\alpha^{2}+4\left(K\chi+j\right)\right]{}^{4}}\right),\label{eq:50}
\end{equation}
}where $\left|\mathcal{C}_{3}\right|$ is an integration constant.
The complete eigenstates are given by{\scriptsize
\begin{equation}
\psi\left(t,r,\varphi,z\right)=\left|\mathcal{C}_{3}\right|\left(\begin{array}{c}
1+\frac{E}{\mathcal{N}}\\
1-\frac{E}{\mathcal{N}}
\end{array}\right)r^{-\frac{1}{2}}\mathrm{e}^{-\left(\sqrt{-E^{2}+K^{2}+m^{2}}\right)r}\left(2r\left(\sqrt{-E^{2}+K^{2}+m^{2}}\right)\right)^{^{\frac{(-2\lambda^{2}+1)\alpha^{2}+2(K\chi+j)^{2}}{2\alpha^{2}}}}F\left(r\right)\label{eq:51}
\end{equation}
}From now on we proceed to study the Feshbach–Villars Oscillator in
a rotating cosmic string space-time . Firstly, we start by considering
a scalar quantum particle embedded in the background gravitational
field of the space-time described by the metric \eqref{eq:42}. In
this way, we shall introduce a replacement of the momentum operator
$p_{i}\longrightarrow p_{i}+im\omega x_{i}$ where $p_{i}=i\nabla_{i}$
in Eq. \eqref{eq:57}. Then, we have
\begin{equation}
\left[\frac{d^{2}}{dr^{2}}+\frac{1}{r}\frac{d}{dr}-m^{2}\omega^{2}r^{2}-\left(\frac{\left(\frac{j+\chi K}{\alpha}\right)^{2}-\lambda^{2}}{r^{2}}\right)-\frac{2\lambda E}{r}+E^{2}-m^{2}-K^{2}+2mw\right]\varphi\left(r\right)=0,\label{eq:52}
\end{equation}
Based on the prior studies, we will apply the GFVT to the case of
KGO in the relevant space using the same techniques as previously.
Inserting Eqs. \eqref{eq:52} and \eqref{eq:45} into the Hamiltonian
\eqref{eq:56}, then assuming the solution \eqref{eq:48}, yields
two linked differential equations comparable to Eq.\eqref{eq:56},
but with different values of $\mathcal{T}^{\prime}$.

Manipulating exactly the same steps before, we obtain the following
radial equation
\begin{equation}
\left[\frac{\partial^{2}}{\partial r^{2}}+\frac{1}{r}\frac{\partial}{\partial r}-m^{2}\omega^{2}r^{2}-\frac{\vartheta^{2}}{r^{2}}-\frac{2\lambda E}{r}+\delta\right]\varphi\left(r\right)=0,\label{eq:53}
\end{equation}
where we have defined
\begin{equation}
\vartheta^{2}=\left(\frac{aE+j}{\alpha}\right)^{2}-\lambda^{2},\qquad\beta^{2}=E^{2}-m^{2}-K^{2}+2mw.\label{eq:54}
\end{equation}
Let us now conside $\mathcal{K}=\sqrt{m\omega}r$, and therefore rewrite
the radial equation (\eqref{eq:53}) as follows ($\varphi\left(r\right)\rightarrow\varphi\left(\mathcal{\mathcal{K}}\right)$):
\begin{equation}
\left[\frac{d^{2}}{d\mathcal{\mathcal{K}}^{2}}+\frac{1}{\mathcal{\mathcal{K}}}\frac{d}{d\mathcal{\mathcal{K}}}-\frac{\vartheta^{2}}{\mathcal{K}^{2}}-\frac{\delta}{\mathcal{K}}-\mathcal{K}^{2}+\frac{\beta^{2}}{m\omega}\right]\varphi\left(\mathcal{\mathcal{K}}\right)=0,\label{eq:55}
\end{equation}
where we have defined a new parameter
\begin{equation}
\delta=2\lambda\sqrt{\frac{m}{\omega}},\label{eq:56}
\end{equation}
Let us look at the asymptotic behavior of the solutions to Eq. (\eqref{eq:55}),
which are found for $\mathcal{\mathcal{K}}\rightarrow0$ and$\mathcal{\mathcal{K}}\rightarrow\infty$.
The behavior of the potential solutions to Eq. (\eqref{eq:55}) at
$\mathcal{\mathcal{K}}\rightarrow0$ and allows us to define the function
$R(\mathcal{K})$ in terms of an unknown function $\mathcal{O}(\mathcal{K})$
as follows from Refs. {[}44-46{]}:
\begin{equation}
\varphi(\mathcal{K})=\mathcal{K}^{\left|\gamma\right|}e^{-\frac{\mathcal{K}^{2}}{2}}\mathcal{O}\left(\mathcal{K}\right),\label{eq:57}
\end{equation}
Hence, by plugging the radial wave function from Eq. (11) into Eq.
(9), we get
\begin{equation}
\frac{d^{2}\mathcal{O}\left(\mathcal{K}\right)}{d\mathcal{\mathcal{K}}^{2}}+\left[\frac{\left(2\left|\gamma\right|+1\right)}{\mathcal{\mathcal{K}}}-2\mathcal{K}\right]\frac{d\mathcal{O}\left(\mathcal{K}\right)}{d\mathcal{\mathcal{K}}}+\left[\frac{\beta^{2}}{m\omega}-2\left(2\left|\gamma\right|+1\right)-\frac{\delta}{\mathcal{K}}\right]\mathcal{O}\left(\mathcal{K}\right)=0,\label{eq:58}
\end{equation}
The Heun biconfluent equation {[}44, 48-50{]} relates to the second
order differential equation (\eqref{eq:58}), and the function $\mathcal{O}\left(\mathcal{K}\right)$
is the Heun biconfluent function.
\begin{equation}
\mathcal{O}\left(\mathcal{K}\right)=HeunB\left(2\left|\gamma\right|,0,\frac{\beta^{2}}{m\omega},2\delta,\mathcal{K}\right)\label{eq:59}
\end{equation}
To continue our discussion of bound state solutions, let us employ
the Frobenius method\citep{key-81,key-84}. As a result, the solution
to Equation (\eqref{eq:59}) may be expressed as a power series expansion
around the origin:
\begin{equation}
\mathcal{O}\left(\mathcal{K}\right)=\stackrel[j=0]{\infty}{\sum}a_{j}\mathcal{K}^{j}\label{eq:60}
\end{equation}
We find the following recurrence connection by substituting the series
(\eqref{eq:60}) into (\eqref{eq:59}):
\begin{equation}
a_{j+2}=\frac{\delta}{\left(j+2\right)\left(j+1+\zeta\right)}a_{j+1}-\frac{\left(\varTheta-2j\right)}{\left(j+2\right)\left(j+1+\zeta\right)}a_{j}\label{eq:61}
\end{equation}
where $\zeta=2\left|\gamma\right|+1$ and $\varTheta=\frac{\beta^{2}}{m\omega}-2\left(\left|\gamma\right|+1\right)$.
We may determine the additional coefficients of the power series expansion
by starting with $a_{0}=1$ and applying the relation (\eqref{eq:60}).
(\eqref{eq:58}). As an example,
\begin{equation}
a_{1}=\frac{\delta}{\zeta}=2\sqrt{\frac{m}{\omega}}\lambda\left(\frac{1}{2\left|\gamma\right|+1}\right)\label{eq:62}
\end{equation}
\begin{align}
a_{2}= & \frac{\delta^{2}}{2\zeta\left(1+\zeta\right)}-\frac{\varTheta}{2\left(1+\zeta\right)}\nonumber \\
= & \left(\frac{m}{\omega}\right)\lambda^{2}\frac{1}{\left(2\left|\gamma\right|+1\right)\left(\left|\gamma\right|+1\right)}-\frac{\varTheta}{4\left(\left|\gamma\right|+1\right)}\label{eq:63}
\end{align}
The wave function must be normalizable, as is widely known in quantum
theory. As a result, we suppose that the function $\mathcal{O}\left(\mathcal{K}\right)$
disappears at $\mathcal{\mathcal{K}}\rightarrow0$ and $\mathcal{\mathcal{K}}\rightarrow\infty$.
This indicates that we have a finite wave function everywhere, which
means that there is no divergence of the wave function at $\mathcal{\mathcal{K}}\rightarrow0$
and $\mathcal{\mathcal{K}}\rightarrow\infty$ thus bound state solutions
may be produced.

In Eq \eqref{eq:58}, we have, however, expressed the function $\mathcal{O}\left(\mathcal{K}\right)$
as a power series expansion around the origin (64). By demanding that
the power series expansion \eqref{eq:60} or the Heun biconfluent
series becomes a polynomial of degree $n$, bound state solutions
can be obtained. As a result, we ensure that $\mathcal{O}\left(\mathcal{K}\right)$
behaves as $\mathcal{K^{\left|\gamma\right|}}$ at $\mathcal{K}\rightarrow0$
the origin and disappears at\citep{key-77,key-78} . We can see from
the recurrence relation \eqref{eq:63} that by applying two constraints
\citep{key-76,key-77,key-78,key-79,key-80,key-81,key-82,key-83},
the power series expansion \eqref{eq:60} becomes a polynomial of
degree $n$:
\begin{equation}
\varTheta=2n\,\,and\,\,a_{n+1}=0\label{eq:64}
\end{equation}
where $n=1,2,3,....$ From the condition $\varTheta=2n$, we can obtain:
\begin{equation}
E^{\pm}\left(n\right)=\pm\sqrt{2m\left(n+\left|\gamma\right|\right)+m^{2}+K^{2}}\label{eq:65}
\end{equation}
The corresponding wave function is given by
\begin{equation}
\varphi\left(r\right)=\left|\mathcal{C}_{4}\right|\mathcal{K}^{\left|\gamma\right|}e^{-\frac{\mathcal{K}^{2}}{2}}HeunB\left(2\left|\gamma\right|,0,\frac{\beta^{2}}{m\omega},2\delta,\mathcal{K}\right)\label{eq:66}
\end{equation}
 the final expression of the wave-function of the spinless FVO propagating
in the dislocation cosmic background can be represented as
\begin{equation}
\psi\left(t,r,\varphi,z\right)=\left|\mathcal{C}_{4}\right|\left(\begin{array}{c}
1+\frac{E}{\mathcal{N}}\\
1-\frac{E}{\mathcal{N}}
\end{array}\right)e^{-i\left(Et-j\varphi-Kz\right)}\mathcal{K}^{\left|\gamma\right|}e^{-\frac{\mathcal{K}^{2}}{2}}HeunB\left(2\left|\gamma\right|,0,\frac{\beta^{2}}{m\omega},2\delta,\mathcal{K}\right),\label{eq:67}
\end{equation}
where the parameters $\vartheta$ and $\delta$ are defined in Eq.\eqref{eq:54}.

\section{Conclusion }

The goal of this work is to investigate the relativistic dynamics
of spinless quantum particles using the Feshbach-Villars representation
of two models, namely the interaction of KGO with the gravitational
field created by the background geometry of: a) cosmic dislocation
and b) cosmic dislocation with Coulomb-Type Potential. We generated
the equivalent formulations in two distinct curved manifolds by modifying
the FV formulation of scalar fields in Minkowski space-time.

We obtained the exact solutions of both systems and we presented the
quantized energy spectra which depend on the parameters that characterize
the space-time topology. 

It is not surprising that the wave-functions of our quantum system
are expressed in terms of the confluent hypergeometric functions for
the free Feshbach-Villars equation in cosmic dilocation space-time,
because the former can be described throughout the latter by using
the appropriate coordinate transformation.

It is worth noting that the Feshbach-Villars oscillator has been presented
under Coulomb-type potential. Nevertheless, as explained in Ref. \citep{key-46},
the electromagnetic 4-vector potential may be introduced using the
same technique by altering the momentum operator as $p_{\alpha}\rightarrow p_{\alpha}-qA_{\alpha}\left(x\right)$.
New and fascinating results linked with the Feshbach-Villars oscillator
can be achieved by dealing with a Coulomb-type potential via a minimum
coupling. Moreover, the wave-functions of our quantum system are represented
in terms of the Biconfluent Heun functions. In this section, we solve
the differential equation using the power series approach.


\begin{thebibliography}{99}
\bibitem{key-1}A. Einstein, Annalen Phys. 49, 769 (1916).

\bibitem{key-2}B. P. Abbott et al., Phys Rev Lett 116, 061102 (2016).

\bibitem{key-3}K. Akiyama et al., Astrophys J Lett 875, L1 (2019).

\bibitem{key-4}R. P. Feynman and A. R. Hibbs, Quantum mechanics and
path integrals, 1965.

\bibitem{key-5}M. D. Schwartz, Quantum field theory and the standard
model, 2013.

\bibitem{key-6}A. Ashtekar and J. J. Stachel, Conceptual problems
of quantum gravity, 1991.

\bibitem{key-7}L. Smolin, The trouble with physics : The rise of
string theory, the fall of a science, and what comes next, 2006.

\bibitem{key-8}N. D. Birrell and P. Davies, Quantum fields in curved
space, 1980.

\bibitem{key-9}L. Parker and D. J. Toms, Quantum field theory in
curved spacetime : Quantized fields and gravity, 2009.

\bibitem{key-10}S. W. Hawking, Comm. Math. Phys. 43, 199 (1975).

\bibitem{key-11}W. G. Unruh and R. M. Wald, Phys. Rev. D 25, 942
(1982).

\bibitem{key-12}G. L. Sewell, Ann. Physics 141, 201 (1982).

\bibitem{key-13}T. W. B. Kibble, J. Phys. A 9, 1387 (1976).

\bibitem{key-14}Y. B. Zel’dovich, Mon. Not. R. Astron Soc. 192, 663
(1980).

\bibitem{key-15}A. Vilenkin, Phys. Rep. 121, 263 (1985).

\bibitem{key-16}T. W. B. Kibble, Phys. Rep. 67, 183 (1980).

\bibitem{key-17}A. Vilenkin, Phys. Lett. B 133, 177 (1983).

\bibitem{key-18}A. Vilenkin and E. P. S. Shellard, Cosmic strings
and other topological defects, 1985.

\bibitem{key-19}M. Moshinsky and Y. F. Smirnov, The harmonic oscillator
in modern physics, 1996.

\bibitem{key-20}A. Ushveridze, Quasi-exactly solvable models in quantum
mechanics, 1994.

\bibitem{key-21}D. Itô, K. Mori, and E. W. Carriere, Il Nuovo Cimento
A (1965-1970) 51, 1119 (1967).

\bibitem{key-22}M. Moshinsky and A. P. Szczepaniak, J. Phys. A 22
(1989).

\bibitem{key-23}S. A. Bruce and P. C. Minning, Il Nuovo Cimento A
(1965-1970) 106, 711 (1993).

\bibitem{key-24}V. V. Dvoeglazov, Il Nuovo Cimento A (1965-1970)
107, 1785 (1994).

\bibitem{key-25}J. Carvalho, A. M. de M. Carvalho, E. Cavalcante,
and C. Furtado, Eur. Phys. J. C 76, 1 (2016).

\bibitem{key-26}L. C. dos Santos and C. de Camargo Barros, Eur. Phys.
J. C 78, 1 (2017).

\bibitem{key-27}R. L. L. Vitória and K. Bakke, Eur. Phys. J. C 78,
1 (2018).

\bibitem{key-28}R. R. Cuzinatto, M. de Montigny, and P. Pompeia,
Class. Quan. Grav 39 (2022).

\bibitem{key-29}F. Ahmed, Europhys. Lett. 131 (2020).

\bibitem{key-30}K. M. Case, Phys Rev 95, 1323 (1954).

\bibitem{key-31}L. L. Foldy, Phys Rev 102, 568 (1956).

\bibitem{key-32}L. L. Foldy and S. A. Wouthuysen, Phys Rev 78, 29
(1950).

\bibitem{key-33}H. Feshbach and F. M. H. Villars, Rev. Modern Phys.
30, 24 (1958).

\bibitem{key-34}B. A. Robson and D. S. Staudte, J. Phys. A : Math.
Gen 29, 157 (1996).

\bibitem{key-35}D. S. Staudte, J. Phys. A 29, 169 (1996).

\bibitem{key-36}M. Merad, L. Chetouani, and A. Bounames, Phys. Lett.
A 267, 225 (2000).

\bibitem{key-37}A. Bounames and L. Chetouani, Phys. Lett. A 279,
139 (2001).

\bibitem{key-38}S. Haouat and L. Chetouani, Eur. Phys. J. C 41, 297
(2005).

\bibitem{key-39}N. Brown, Z. Papp, and R. M. Woodhouse, Few-Body
Systems 57, 103 (2015).

\bibitem{key-40}B. Motamedi, T. Shannon, and Z. Papp, Few-Body Systems
(2019).

\bibitem{key-41}Bouzenada A, Boumali A. arXiv preprint arXiv:2302.02268.
2023.

\bibitem{key-42}Q. Wen-Chao, Chinese Phys. 12, 1054 (2003).

\bibitem{key-43}H. Motavalli and A. R. Akbarieh, Mod. Phys. Lett.
A 25, 2523 (2010).

\bibitem{key-44}F. Yasuk, A. Durmus and I. Boztosun, J. Math. Phys
47, 082302 (2006).

\bibitem{key-45}A. L. Cavalcanti de Oliveira and E. R. Bezerra de
Mello, Class. Quantum Grav. 23, 5249 (2006).

\bibitem{key-46}W. Greiner, Relativistic Quantum Mechanics: Wave
Equations, 3rd Edition (Springer, Berlin, 2000).

\bibitem{key-47}F. L. Gross, Relativistic quantum mechanics and field
theory, 1993.

\bibitem{key-48}O. Klein, Z. Phys 37, 895 (1926).

\bibitem{key-49}W. Gordon, Z. Phys 40, 117 (1926).

\bibitem{key-50}F. L. Gross, Relativistic quantum mechanics and field
theory, 1993.

\bibitem{key-51}A. J. Silenko, Phys. Rev. A 77, 012116 (2008).

\bibitem{key-52}Vitória, R. L. L., and K. Bakke. EPJC 78,1-6,(2018).

\bibitem{key-53}Hiscock, Phys. Rev. D 31 12, 3288 (1985).

\bibitem{key-54}I. Gott, J. R., Astrophys J 288, 422 (1985).

\bibitem{key-55}Bakke, Knut, et al. Phys. Rev. D 79.2 (2009).

\bibitem{key-56}J. Carvalho, A. M. de M. Carvalho, E. Cavalcante,
and C. Furtado, Eur. Phys. J. C 76, 1 (2016).

\bibitem{key-57}A. Boumali and N. Messai, Can. J. Phys. 92, 1460
(2014).

\bibitem{key-58}A. J. Silenko, Theoret. Math. Phys. 156, 1308 (2008).

\bibitem{key-59}A. J. Silenko, Phys. Rev. D 88, 045004 (2013).

\bibitem{key-60}A. Mostafazadeh, J. Phys. A 31, 7829 (1998).

\bibitem{key-61}M. Abramowitz and I. A. Stegun, Handbook of mathematical
functions with formulas, graphs, and mathematical tables, volume 55,
Dover Publications, New York, 1970.

\bibitem{key-62}E. Bragança, H. S. Mota, and E. B. de Mello, Int
J Mod Phys D 24, 1550055 (2015).

\bibitem{key-63}F. Ahmed, Sci Rep 12 (2022).

\bibitem{key-64}G. Arfken, H. Weber, and F. Harris, Mathematical
Methods for Physicists : A Comprehensive Guide, Elsevier Science,
2012.

\bibitem{key-65}Mazur, Phys. Rev. Lett 57 8, 929 (1986).

\bibitem{key-66}P. S. Gerbert and R. Jackiw, Comm. Math. Phys. 124,
229 (1989).

\bibitem{key-67}Matsas, Phys. Rev. D 42 8, 2927 (1990).

\bibitem{key-68}Bekenstein, Phys. Rev. D 45 8, 2794 (1992).

\bibitem{key-69}J. S. Anandan, Phys. Lett. A 195, 284 (1994).

\bibitem{key-70}V. De Lorenci and E. Moreira, Phys. Rev. D 63, 027501
(2000).

\bibitem{key-71}V. De Lorenci and E. Moreira, Nuclear Phys. B Proc.
Suppl. 127, 150 (2004).

\bibitem{key-72}M. S. Cunha, C. R. Muniz, H. R. Christiansen, and
V. B. Bezerra, Eur. Phys. J. C 76, 1 (2016).

\bibitem{key-73}B.-Q. Wang, Z. W. Long, C.-Y. Long, and S. Wu, Modern
Phys. Lett. A 33, 1850025 (2018).

\bibitem{key-74}Z. Wang, Z. W. Long, C.-Y. Long, and B.-Q. Wang,
Can. J. Phys. 95, 331 (2017).

\bibitem{key-75}M. Hosseinpour, H. Hassanabadi, and M. de Montigny,
Int. J. Geom. Methods Mod. Phys. (2018).

\bibitem{key-76}F. Ahmed, Europhys. Lett. 130 (2020).

\bibitem{key-77}A. Verćin, Phys. Lett. B 260, 120 (1991).

\bibitem{key-78}J. Myrhein, E. Halvorsen and A. Verćin, Phys. Lett.
B 278, 171 (1992).

\bibitem{key-79}K. Bakke and F. Moraes, Phys. Lett. A 376, 2838 (2012).

\bibitem{key-80}K. Bakke, Ann. Phys. (NY) 341, 86 (2014).

\bibitem{key-81}C. Furtado, B. G. C. da Cunha, F. Moraes, E. R. Bezerra
de Mello and V. B. Bezerra, Phys. Lett. A 195, 90 (1994).

\bibitem{key-82}K. Bakke and H. Belich, Eur. Phys. J. Plus 127, 102
(2012).

\bibitem{key-83}K. Bakke and H. Belich, Ann. Phys. (NY) 333, 272
(2013).

\bibitem{key-84}G. B. Arfken and H. J. Weber, Mathematical Methods
for Physicists, sixth edition (Elsevier Academic Press, New York,
2005).

\end{thebibliography}
\end{document}